\documentclass[conference]{IEEEtran}

\usepackage[T1]{fontenc}
\usepackage[utf8]{inputenc}
\usepackage[british]{babel}

\usepackage{amsmath,amssymb,amsfonts}
\usepackage{mathtools}
\usepackage{bm}

\usepackage{tikz}
\usetikzlibrary{arrows.meta,calc,positioning}

\usepackage{graphicx}
\usepackage{cite}
\usepackage{tabularx}
\usepackage{array}
\newcolumntype{Y}{>{\raggedright\arraybackslash}X}

\usepackage{xcolor}
\usepackage[caption=false,font=footnotesize]{subfig} 
\usepackage[
  unicode=true,
  pdfencoding=auto,
  psdextra,
  hidelinks
]{hyperref}

\usepackage{microtype}
\title{Hagenberg Risk Management Process (Part 1): Multidimensional Polar Heatmaps for Context-Sensitive Risk Analysis}

\author{\IEEEauthorblockN{Eckehard Hermann,
Harald Lampesberger}
\IEEEauthorblockA{Department Secure Information Systems, 
University of Applied Sciences Upper Austria\\
Email: eckehard.hermann@fh-hagenberg.at,
harald.lampesberger@fh-hagenberg.at}}

\begin{document}
\maketitle

\begin{abstract}
Traditional two-dimensional risk matrices (heatmaps) are widely used to model and visualize likelihood and impact relationships, but they face fundamental methodological limitations when applied to complex infrastructures.
In particular, regulatory frameworks such as NIS2 and DORA call for more context-sensitive and system-oriented risk analysis.
We argue that incorporating contextual dimensions into heatmaps enhances their analytical value.
As a first step towards our \textit{Hagenberg Risk Management Process} for complex infrastructures and systems, this paper introduces a multidimensional (ND) polar heatmap as a formal model that explicitly integrates additional context dimensions and subsumes classical two-dimensional models as a special case.
\end{abstract}

\section{Introduction and problem statement}

The reliable operation of digital infrastructures has become a basic
prerequisite of modern societies.
Data centers form the technical backbone for critical services in business,
public administration and finance
\cite{JRCDataCentersEnergy2024}.
Their spatial and environmental impacts are also increasingly in focus \cite{MonstadtSaltzman2025}.
With growing interconnection, automation and power density, however, both the
technical complexity and the dependence of core societal functions on the
uninterrupted availability of these systems are increasing.

In parallel, regulatory frameworks such as the EU NIS2 Directive \cite{NIS2} and
the Digital Operational Resilience Act (DORA) \cite{DORA}
significantly tighten the requirements for risk management.
Operators are required to identify, assess and treat risks not merely in
isolation, but holistically, context-dependently and in a traceable manner
\cite{NIS2}.
In particular, the focus shifts to dependencies, system states, operating modes, and resilience mechanisms.

Standards, such as ISO/IEC 31010 \cite{ISO31010}, provide a wide range of risk assessment instruments, however, in operational practice, the risk analysis of technical systems is still often
performed using two-dimensional risk matrices
\cite{Duijm2015RiskMatrices}.
These so-called heatmaps represent risks as a function of two dimensions,
typically likelihood of occurrence and impact.
Owing to their ease of use and strong communicative effect, they are firmly
established in many organisations and are implicitly assumed in standards and
guidance
\cite{ISO31010}.

At the same time, experience from research and practice is accumulating that this
form of risk mapping has significant limitations in complex, highly
available infrastructures because of its lack of context.
Modern technical systems are characterised by a multitude of context-dependent
influencing factors, including redundancy concepts, maintenance status,
environmental conditions, load profiles, and operating strategies.
These factors affect not only likelihoods and impacts, but also their
interactions.
ISO/IEC 31010~\cite{ISO31010} denotes instruments that allow modelling context, but they require substantially more effort, particularly when used for risk triage.

The problem addressed in this paper is how risks can be modelled and visualized by extending the well established heatmap concept so that additional contextual dimensions are considered explicitly, without sacrificing comprehensibility or compatibility with existing risk management processes.
Against this background, we propose a multidimensional (ND) polar heatmap that
subsumes classical 2D risk matrices as a special case while enabling a more
context-sensitive analysis.

\section{State of the Art}
\label{sec:soa}

Risk analysis of technical and socio-technical systems has for decades been a
core element of safety, reliability, and resilience management.
Depending on the application domain, system complexity and regulatory context,
different methodological approaches have become established, each addressing
specific objectives, levels of abstraction and decision support.
While early approaches mainly aimed at qualitative classification, today a
broad spectrum of qualitative, semi-quantitative and quantitative methods
exists.

The basic motivation and problem statement were outlined in Section~I.
This section systematically categorises the relevant approaches and discusses
their suitability with regard to context-sensitive, multidimensional risk
modelling as addressed by ND polar heatmaps.

\subsection{Qualitative and semi-quantitative risk matrices}

Qualitative and semi-quantitative risk matrices remain the most widely used
approach in practical risk analysis.
Risks are classified along two dimensions—typically likelihood of occurrence
and impact—into discrete classes and visualized as a heatmap.

International standards such as ISO~31000 \cite{ISO31000} and ISO/IEC~31010 \cite{ISO31010}
treat risk matrices as an established instrument for risk identification,
assessment and communication, particularly in early phases of risk management.
Formally, a classical risk matrix can be modelled as a mapping
\begin{equation}
R = f(P, I)
\end{equation}
where probability $P$ and impact $I$ denote discretized, ordinally scaled quantities.

The main advantage of this approach is its low barrier to entry, intuitive
interpretability, and broad acceptance in management and audit contexts.
Risk matrices enable rapid prioritisation \cite{Duijm2015RiskMatrices}.
At the same time, they are often used as a means of communication between
technical and non-technical stakeholders \cite{Sutherland2021RiskMatrices}.

\subsection{Methodological limitations of classical risk matrices}

Despite their widespread use, the theoretical and practical limitations of
risk matrices have long been the subject of scientific criticism.
Kaplan and Garrick~\cite{KaplanGarrick1981} argue fundamentally that risk should not be understood as a
scalar quantity, but as a set of possible scenarios, each with different
probabilities and consequences.

Cox~\cite{Cox2008} shows that discrete risk matrices can, under certain conditions, produce
systematic ranking errors, such that risks with higher expected loss are rated
lower than risks with lower loss.
These effects result from non-linear class boundaries, subjectivity of the
scales, and the implicit assumption that likelihood and impact are
independent.

Moreover, risk matrices largely abstract away from the current system state.
Context factors such as redundancy, operating mode, maintenance status, or
environmental influences are either not considered at all or only implicitly.
Classical heatmaps are therefore only of limited suitability for assessing
complex, dynamic systems.

\subsection{FMEA and FMECA}

Failure Modes and Effects Analysis (FMEA) and its extended form FMECA, incorporating criticality, are
established methods for the systematic identification of failure modes, their
causes, and their effects.
In standardised form (e.\,g.\ IEC~60812), failure modes are assessed using
discrete criteria such as severity, occurrence, and detection \cite{IEC60812}.

The strength of FMEA/FMECA lies in structured, component-level analysis and its
tight coupling to engineering processes.
However, multidimensional context states are primarily modelled in tabular
form.
A consistent, visually integrated representation of complex state spaces is
not envisaged.
In contrast, the ND polar heatmap models context dimensions explicitly as axes
of a formal state space.

\subsection{HAZOP}

Hazard and Operability Studies (HAZOP) are guideword-based, interdisciplinary
methods for identifying deviations from design intent, particularly in
process-oriented systems \cite{IEC61882}.
HAZOP is well suited to the systematic elicitation of potential hazards.

As a standalone risk-modelling approach, however, HAZOP does not provide a
compact aggregation or classification across multiple context dimensions.
Results are typically produced as qualitative lists of scenarios.
ND polar heatmaps can absorb and structure these results without replacing
HAZOP.

\subsection{LOPA}

Layer of Protection Analysis (LOPA) is a semi-quantitative approach for
assessing risk scenarios while taking independent protection layers into
account \cite{CCPSLOPA}.
The focus lies on reasoning about barriers and their risk-reduction effect.

LOPA is strongly scenario-oriented and primarily addresses linear
cause–effect chains.
A simultaneous representation of many context states in a multidimensional
structure is not envisaged.
In contrast, the ND polar heatmap targets a state-based modelling approach that
is compatible with rule representations.

\subsection{Standards-based IT and cyber risk models}

Standards such as ISO/IEC~27005 \cite{ISO27005} and NIST~SP~800-30 \cite{NIST80030}
define structured processes for identifying, assessing, and treating
information-security risks.
The emphasis is on methodological process steps, not on the concrete
representation form of the results.

In practice, the results are often transferred back into classical risk
matrices or risk registers.
ND polar heatmaps can serve as a formal representation and visualization layer
that maps standards-compliant assessments into multidimensional state spaces
with less loss of information.

\subsection{FAIR}

Factor Analysis of Information Risk (FAIR) is a quantitative model that defines
risk as a function of loss event frequency and loss magnitude and further
decomposes both \cite{FAIRBook}.
FAIR supports precise, data-driven risk reasoning.

However, FAIR is primarily numerical and does not provide an integrated visual
ND state representation.
The ND polar heatmap is therefore best understood not as a replacement, but as
a complementary, communication-friendly structure.

\subsection{STPA}

Systems-Theoretic Process Analysis (STPA) is based on a systems-theoretic view
of safety and analyses hazards via control structures and unsafe control
actions \cite{Leveson2011}.
The approach is particularly suitable for software-intensive and adaptive
systems.

STPA, however, focuses on causality and control logic rather than on a
multidimensional classification of risk and context states.
An ND heatmap representation is not part of the approach.

\subsection{Summary and positioning with respect to ND polar heatmaps}

In summary, existing risk methods are either strongly qualitative,
component- or scenario-oriented, or primarily quantitative.
None of the established methods provides a formal, state-based and at the same
time visually integrated mapping of multidimensional context factors.

The ND polar heatmap therefore positions itself as a representation and
integration layer that does not replace existing methods, but consolidates
their results in a consistent, regulatorily compatible, and decision-oriented
manner.

To systematically position the approaches discussed above,
Table~\ref{tab:riskmethods} summarizes their key characteristics, strengths and
limitations.
The comparison follows criteria relevant to regulatory and context-sensitive
analyses, in particular dimensionality, context representation, visualization
and suitability for management use.

\begin{table*}[t]
\centering
\caption{Comparison of established risk-modelling approaches in the context of
multidimensional and regulatory requirements}
\label{tab:riskmethods}
\setlength{\tabcolsep}{4pt}
\renewcommand{\arraystretch}{1.2}
\small
\begin{tabularx}{\textwidth}{|
p{2.6cm}|
p{1.8cm}|
p{2.2cm}|
p{2.2cm}|
Y|
Y|}
\hline
\textbf{Method} &
\textbf{Dimen-\newline sionality} &
\textbf{Context-\newline representation} &
\textbf{Visuali-\newline sation} &
\textbf{Strengths} &
\textbf{Key limitations compared with the ND polar heatmap} \\
\hline

2D risk matrix (heatmap) &
2D (Likelihood, Impact) &
implicit / heavily reduced &
very good, intuitive &
Simple, widely used, suitable for management &
Loss of contextual information; ranking errors; no explicit modelling of
redundancy, operating states or dependencies \\
\hline

FMEA / FMECA &
multiple attributes, tabular &
indirect (attributes) &
limited &
Structured component-level failure analysis &
No integrated ND state representation; limited aggregation and visualization
capability \\
\hline

HAZOP &
qualitative, guideword-based &
explicit (scenarios) &
none &
Systematic identification of deviations &
No formal risk aggregation; no overall visual overview; not comparative \\
\hline

LOPA &
scenario-based &
focused on protection layers &
limited &
Transparent barrier reasoning &
Not state-based; unsuitable for parallel context dimensions \\
\hline

Quantitative risk analysis (QRA) &
high (continuous) &
explicit, model-based &
limited &
High analytical precision &
High data and modelling effort; limited communicability \\
\hline

FAIR &
quantitative (frequency, loss) &
parametric &
limited &
Economic view of risk &
No integrated visual ND representation; complex for non-technical
stakeholders \\
\hline

STPA &
systems-theoretic &
explicit (control structures) &
limited &
Captures systemic causes &
No risk scoring; no ND state classification or heatmap representation \\
\hline

\textbf{ND polar heatmap (this paper)} &
\textbf{arbitrary (ND)} &
\textbf{explicit as axes} &
\textbf{high, integrated} &
\textbf{Formal ND mapping, context-sensitive, compatible with regulatory requirements} &
\textbf{Requires initial axis definition and rule modelling} \\
\hline

\end{tabularx}
\end{table*}
\subsection{Positioning the ND polar heatmap as a formal representation layer}

Multidimensional polar heatmaps should not be understood as competing with
established risk-management methods such as FMEA/FMECA, HAZOP, LOPA, FAIR, or
standards-based IT risk approaches.
Rather, they address an aspect that is often only insufficiently formalized in
these methods: the consistent, multidimensional \emph{representation and
aggregation of risk and context states}.
While the methods above primarily focus on structured elicitation of hazards,
scenarios, cause--effect chains, or quantitative loss models, the question of an
overarching, comparable state representation often remains implicit or is
reduced to two-dimensional projections.

Heatmaps in polar or radial coordinates (\emph{polar/radial heatmaps}) are a
well-established visualization for cyclic, direction-dependent or
longitudinally periodic data and are, for example, used to depict longitudinal
time series (e.\,g.\ in actigraphy) \cite{Keogh2020RadialHeatmap}.
Circular heatmap tracks and related radial representations are also supported
by established visualization tools \cite{Krzywinski2009Circos}.
In the context of multivariate profiles, radial heat maps are additionally used
as a comparative representation \cite{Kremer2024RadialHeatmaps}.
Circular visualization approaches such as \textit{Circos} or \textit{circlize} implement annular
heatmap ``tracks'' to stack multidimensional data along a circular axis system
\cite{Krzywinski2009Circos}.
At the tooling level, the R package \texttt{circlize} provides a flexible
implementation of circular layouts \cite{Gu2014circlize}.
As an example of a high-performance, web-based implementation of declarative
chart types, \textit{ECharts} describes an extensible framework for interactive
visualizations \cite{VegaLite2017}.
This paper transfers this visual principle into a formal, multidimensional risk
model with explicit axis and aggregation semantics and a traceable mapping
function $H:\mathcal{L}\rightarrow G$.

The ND polar heatmap therefore positions itself as a formal \emph{mapping layer}
on top of existing approaches.
Results from FMEA, HAZOP, STPA, or FAIR can be integrated into the state space
$\mathcal{L}$ as discrete context dimensions, states or rule assignments,
without replacing or diluting their respective methodological strengths.
Unlike tabular or purely narrative representations, the ND polar heatmap enables
an explicit, rule-based mapping $H:\mathcal{L}\rightarrow G$ that is both
systematically comparable and visually interpretable.

Thus, the ND polar heatmap acts as a connecting element between
detail-oriented risk analysis and management-friendly, regulatorily compatible
decision support.
It allows heterogeneous risk sources, context factors and analysis results to
be consolidated within a shared multidimensional model space, without reducing
them to a purely qualitative or two-dimensional view.
In particular, in the context of regulatory requirements, such as NIS2 and DORA,
this representation perspective provides substantial added value because it does
not replace existing methods but integrates them structurally and makes them
comparable.

\section{Formal definition of an ND polar heatmap}
\label{sec:ndpolarformal}

\subsection{Axes, levels and state space}

Let $d \in \mathbb{N}$ be the number of risk dimensions (axes).
We model each axis $i \in \{1,\dots,d\}$ by a finite, ordered set of discrete
levels
\begin{equation}
L_i \coloneqq \{0,1,\dots,n_i-1\},
\end{equation}
where $n_i \ge 2$ denotes the number of levels on axis $i$.
The ordering corresponds to increasing intensity (e.\,g.\ ``low'' $\to$ ``high'').

The \emph{ND state space} (discrete context state) is the Cartesian product
\begin{equation}
\mathcal{L} \coloneqq \prod_{i=1}^{d} L_i.
\end{equation}
A state $\bm{\ell} \in \mathcal{L}$ is a $d$-tuple
\begin{equation}
\bm{\ell} = (\ell_1,\ell_2,\dots,\ell_d), \quad \ell_i \in L_i.
\end{equation}

\subsection{Risk classes and mapping function}

Let $G$ be a finite set of risk classes (\textit{grades}) with a total order $\preceq$,
e.\,g.\ $G=\{\text{very-low},\text{low},\text{moderate},\text{high},\text{critical}\}$.
Formally, an \emph{ND heatmap} is a mapping function
\begin{equation}
H:\mathcal{L}\rightarrow G.
\label{eq:Hdef}
\end{equation}
Thus, each context state $\bm{\ell}$ is assigned exactly one grade
$g = H(\bm{\ell})$.

\paragraph*{Remark (2D as a special case).}
For $d=2$, we obtain the classical risk matrix
\begin{equation}
H_{2D}: L_1 \times L_2 \rightarrow G,
\end{equation}
typically with $L_1$ representing likelihood levels and $L_2$ representing
impact levels.
The transition to $d>2$ models additional context factors explicitly (e.\,g.\
redundancy level, operating mode, maintenance status, ambient temperature).

\subsection{Polar embedding: angle, radius and sector geometry}

The polar visualization assigns each axis $i$ an angular range (sector).
For an even angular partition we define the sector width
\begin{equation}
\Delta\theta \coloneqq \frac{2\pi}{d},
\end{equation}
and the sector center for axis $i$ as
\begin{equation}
\theta_i \coloneqq \theta_0 + \left(i-\frac{1}{2}\right)\Delta\theta,
\label{eq:theta_i}
\end{equation}
where $\theta_0$ is a freely chosen rotation constant.

For radial scaling, the discrete level $\ell_i\in\{0,\dots,n_i-1\}$ is mapped to
a normalised radius value:
\begin{equation}
\rho_i(\ell_i) \coloneqq \frac{\ell_i + \tfrac{1}{2}}{n_i} \in \left(0,1\right).
\label{eq:rho}
\end{equation}

\paragraph*{Geometric primitive (annular sector).}
For each axis $i$ and level $\ell_i$ there is an annular segment
\begin{equation}
\begin{aligned}
S_{i,\ell_i} \coloneqq \Big\{(r,\theta)\ \Big|\ &\theta \in [\theta_0+(i-1)\Delta\theta,\ \theta_0+i\Delta\theta),\\
&r \in \Big[\frac{\ell_i}{n_i},\ \frac{\ell_i+1}{n_i}\Big)\Big\}.
\end{aligned}
\label{eq:sector}
\end{equation}

For $d>2$, the ND visualization is obtained as a \emph{slice} through the ND
state space by fixing all axes $k\ge 3$ to slice indices $\sigma_k \in L_k$.

\subsection{ND slice definition and compatible rule representation}

Let $\bm{\sigma}=(\sigma_3,\dots,\sigma_d)$ be a slice vector for axes
$3,\dots,d$.
The slice defines a 2D partial function
\begin{equation}
H_{\bm{\sigma}}: L_1\times L_2 \rightarrow G,\qquad
H_{\bm{\sigma}}(\ell_1,\ell_2) \coloneqq H(\ell_1,\ell_2,\sigma_3,\dots,\sigma_d).
\label{eq:slice}
\end{equation}

\subsection{Acceptance thresholds}

\begin{figure}[t]
\centering
\begin{tikzpicture}[scale=1.0, line cap=round, line join=round]
\def\R{2.6}
\def\step{120}
\def\start{-90}

\foreach \k in {1,2,3,4}{
  \draw[black!25] (0,0) circle ({\R*\k/4});
}
\foreach \i in {0,1,2}{
  \draw[black!35] (0,0) -- ({\R*cos(\start+\i*\step)},{\R*sin(\start+\i*\step)});
}
\draw[black!50] (0,0) circle (\R);

\fill[green!35] (0,0)
  -- ({\R*1/4*cos(\start)},{\R*1/4*sin(\start)})
  arc[start angle=\start, end angle=\start+\step, radius=\R*1/4] -- cycle;

\fill[yellow!45] (0,0)
  -- ({\R*2/4*cos(\start)},{\R*2/4*sin(\start)})
  arc[start angle=\start, end angle=\start+\step, radius=\R*2/4]
  -- ({\R*1/4*cos(\start+\step)},{\R*1/4*sin(\start+\step)})
  arc[start angle=\start+\step, end angle=\start, radius=\R*1/4] -- cycle;

\fill[orange!45] (0,0)
  -- ({\R*3/4*cos(\start)},{\R*3/4*sin(\start)})
  arc[start angle=\start, end angle=\start+\step, radius=\R*3/4]
  -- ({\R*2/4*cos(\start+\step)},{\R*2/4*sin(\start+\step)})
  arc[start angle=\start+\step, end angle=\start, radius=\R*2/4] -- cycle;

\fill[green!25] (0,0)
  -- ({\R*2/4*cos(\start+\step)},{\R*2/4*sin(\start+\step)})
  arc[start angle=\start+\step, end angle=\start+2*\step, radius=\R*2/4] -- cycle;

\fill[red!35] (0,0)
  -- ({\R*cos(\start+\step)},{\R*sin(\start+\step)})
  arc[start angle=\start+\step, end angle=\start+2*\step, radius=\R]
  -- ({\R*2/4*cos(\start+2*\step)},{\R*2/4*sin(\start+2*\step)})
  arc[start angle=\start+2*\step, end angle=\start+\step, radius=\R*2/4] -- cycle;

\fill[yellow!35] (0,0)
  -- ({\R*1/4*cos(\start+2*\step)},{\R*1/4*sin(\start+2*\step)})
  arc[start angle=\start+2*\step, end angle=\start+3*\step, radius=\R*1/4] -- cycle;

\fill[orange!35] (0,0)
  -- ({\R*2/4*cos(\start+2*\step)},{\R*2/4*sin(\start+2*\step)})
  arc[start angle=\start+2*\step, end angle=\start+3*\step, radius=\R*2/4]
  -- ({\R*1/4*cos(\start+3*\step)},{\R*1/4*sin(\start+3*\step)})
  arc[start angle=\start+3*\step, end angle=\start+2*\step, radius=\R*1/4] -- cycle;

\draw[black, line width=1.2]
  ({\R*2/4*cos(\start)},{\R*2/4*sin(\start)})
  arc[start angle=\start, end angle=\start+\step, radius=\R*2/4];
\draw[black, line width=1.2]
  ({\R*3/4*cos(\start+\step)},{\R*3/4*sin(\start+\step)})
  arc[start angle=\start+\step, end angle=\start+2*\step, radius=\R*3/4];
\draw[black, line width=1.2]
  ({\R*1/4*cos(\start+2*\step)},{\R*1/4*sin(\start+2*\step)})
  arc[start angle=\start+2*\step, end angle=\start+3*\step, radius=\R*1/4];

\node[font=\footnotesize, align=center]
  at ({(\R+0.45)*cos(\start+\step/2)},{(\R+0.45)*sin(\start+\step/2)})
  {Axis 1\\(e.\,g.\ likelihood)};
\node[font=\footnotesize, align=center]
  at ({(\R+0.45)*cos(\start+3*\step/2)},{(\R+0.45)*sin(\start+3*\step/2)})
  {Axis 2\\(e.\,g.\ impact)};
\node[font=\footnotesize, align=center]
  at ({(\R+0.45)*cos(\start+5*\step/2)},{(\R+0.45)*sin(\start+5*\step/2)})
  {Axis 3\\(context)};

\fill[black] (0,0) circle (0.03);
\end{tikzpicture}
\caption{Stylised ND polar heatmap for $d=3$. Annular sectors encode discrete levels per axis; colors encode risk classes $G$ according to $H$ in \eqref{eq:Hdef}. Black arcs indicate example acceptance thresholds $\bm{a}$.}
\label{fig:ndpolar_tikz}
\end{figure}
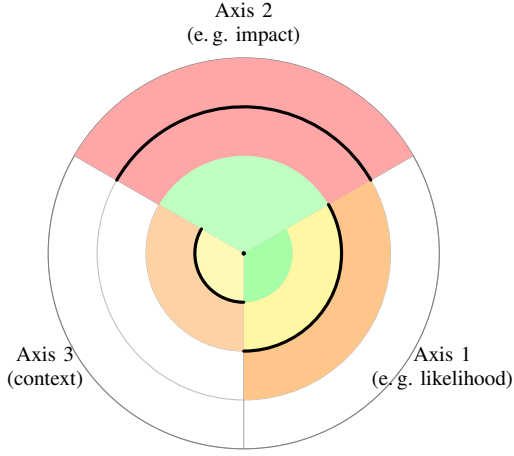

To model per-axis acceptance thresholds we define a threshold vector
$\bm{a}\in\mathcal{L}$,
\begin{equation}
\bm{a}=(a_1,\dots,a_d), \quad a_i\in L_i,
\end{equation}
an axis-violation function
\begin{equation}
v_i(\ell_i) \coloneqq \mathbb{I}[\ell_i > a_i],
\end{equation}
and the aggregated violation count
\begin{equation}
V(\bm{\ell}) \coloneqq \sum_{i=1}^d v_i(\ell_i).
\label{eq:violation}
\end{equation}
In the visualization, $a_i$ can be drawn as a radial arc on axis $i$.
Fig.~\ref{fig:ndpolar_tikz} shows a stylized ND polar heatmap for $d=3$ axes.

\subsection{Summary of the formal core elements}

Formally, the ND polar heatmap is characterised by:
(i) axis level sets $L_i$ and the state space $\mathcal{L}$,
(ii) a mapping $H:\mathcal{L}\rightarrow G$,
(iii) a polar embedding via \eqref{eq:theta_i} and \eqref{eq:rho}, and
(iv) optional acceptance thresholds $\bm{a}$ to visualize threshold violations
via \eqref{eq:violation}.
Thus, $d=2$ is a projection/specialisation, whereas $d>2$ models context
dimensions explicitly with low information loss.

\section{Comparison of 2D risk matrix vs.\ ND polar heatmap using the scenario ``data-center cooling failure''}
\label{sec:comparison_matrix_polar}

This section compares the classical 2D risk matrix (as \emph{context-indexed
slices}) with the ND polar heatmap (as an \emph{aggregating, exploratory
visualization layer}) using the scenario ``data-center cooling failure'' (see
Fig.~\ref{fig:matrix_vs_polar_overview} to Fig.~\ref{fig:matrix_vs_polar_overdue}).
The focus is on analyzing context factors, in particular, the
\emph{cooling redundancy level} with \emph{maintenance status} held constant.

\subsection{Semantics of the 2D matrix as context-indexed slices}
\label{sec:matrix_slices_semantics}

It is important to state the correct semantics of a 2D matrix in an ND context:
entries of a matrix (likelihood $\times$ impact) are \emph{not an aggregation};
instead, they are specified \emph{separately for each context state}.
Formally, let $d \ge 2$ be the number of axes and
$\mathcal{L}=\prod_{i=1}^d L_i$ the discrete state space (cf.\
Section~\ref{sec:ndpolarformal}).
We distinguish two primary axes (here: likelihood $L_1$ and impact $L_2$) and
context axes $L_3,\dots,L_d$.
For each context state $\bm{\sigma}=(\sigma_3,\dots,\sigma_d)\in \prod_{k=3}^d L_k$
there is a separate 2D matrix (a so-called \emph{slice})
\begin{equation}
M_{\bm{\sigma}}: L_1 \times L_2 \rightarrow G,
\label{eq:matrix_slice}
\end{equation}
whose values $M_{\bm{\sigma}}(\ell_1,\ell_2)$ are determined by the analyst or
by rules/heuristics.
The currently viewed slice $M_{\bm{\sigma}}$ is typically configured outside the annular visualization in a classical user interface, e.g.,  via text fields/drop-downs (context layer controls),
while the matrix itself remains unchanged in the 2D plane.

\subsection{Aggregation semantics of the ND polar heatmap for the primary axes}
\label{sec:polar_axis_aggregation}

The ND polar heatmap represents the context axes explicitly as additional axes
(point positions), while the primary axes (likelihood/impact) display
color-coded \emph{aggregated} information derived from the currently viewed
slice $M_{\bm{\sigma}}$.
The color of the likelihood axis for a column $\ell_1\in L_1$ is determined by
the \emph{most frequent} risk class in that column (mode), and analogously for
impact from the corresponding row.
In the event of a tie, the \emph{higher} risk grade is selected.
An additional special rule prioritises the \emph{risk cell} itself in the row/column in which the currently considered risk is located.

\paragraph*{Preparation: ordering and mode with tie-break.}
Let $G$ be a finite, totally ordered set of risk grades with order $\preceq$.
For a multiset $\mathcal{S}\subseteq G$ define
\begin{equation}
\operatorname{mode}_{\preceq}(\mathcal{S})
\;\coloneqq\;
\arg\max_{g\in G}\big|\{s\in \mathcal{S}\mid s=g\}\big|,
\end{equation}
where, if there are multiple maximizers, the \emph{greatest} grade with respect to $\preceq$ is chosen.

\paragraph*{Risk position.}
Let the current risk position in the 2D matrix be given as
\begin{equation}
\bm{r}=(r_1,r_2)\in L_1\times L_2.
\end{equation}
The visualization in a user interface is then a bold frame around the corresponding 2D matrix cell and a cross in the polar heatmap segment.

\paragraph*{Aggregation for likelihood (columns).}
For fixed $\bm{\sigma}$ and a column $\ell_1\in L_1$ we define the
column multiset
\begin{equation}
\mathcal{C}_{\bm{\sigma}}(\ell_1)\;\coloneqq\;\{M_{\bm{\sigma}}(\ell_1,\ell_2)\mid \ell_2\in L_2\}.
\end{equation}
Then the aggregated axis color for the likelihood axis is:
\begin{equation}
A^{(x)}_{\bm{\sigma}}(\ell_1)\;\coloneqq\;
\begin{cases}
M_{\bm{\sigma}}(r_1,r_2), & \text{if } \ell_1=r_1,\\[4pt]
\operatorname{mode}_{\preceq}\big(\mathcal{C}_{\bm{\sigma}}(\ell_1)\big), & \text{otherwise.}
\end{cases}
\label{eq:agg_x}
\end{equation}

\paragraph*{Aggregation for impact (rows).}
Analogously, define the row multiset
\begin{equation}
\mathcal{R}_{\bm{\sigma}}(\ell_2)\;\coloneqq\;\{M_{\bm{\sigma}}(\ell_1,\ell_2)\mid \ell_1\in L_1\},
\end{equation}
and the aggregated axis color for the impact axis:
\begin{equation}
A^{(y)}_{\bm{\sigma}}(\ell_2)\;\coloneqq\;
\begin{cases}
M_{\bm{\sigma}}(r_1,r_2), & \text{if } \ell_2=r_2,\\[4pt]
\operatorname{mode}_{\preceq}\big(\mathcal{R}_{\bm{\sigma}}(\ell_2)\big), & \text{otherwise.}
\end{cases}
\label{eq:agg_y}
\end{equation}

This makes the distinction explicit: the 2D matrix itself is \emph{not}
aggregated (it is the analyst-defined slice $M_{\bm{\sigma}}$), whereas the polar
heatmap shows, for the primary axes, an \emph{aggregation} over rows/columns of
that slice.
The context axes, by contrast, are not aggregated; they are shown explicitly in their coloring in the visualized slice and the context state is denoted by dots.

\subsection{Visual analyzability of context variations}
\label{sec:visual_analytics}

In a 2D matrix, every change of a context layer $\sigma_k$ (e.\,g.\ redundancy
level) has to be understood as switching to a different slice
$M_{\bm{\sigma}}$.
In the matrix view, the current context is typically exposed and configured via a user interface, e.g., text
fields/drop-downs; the matrix itself provides no immediate visual indication of
\emph{which} context state is currently active.

In the polar heatmap, context states are visually coupled:
(i) the cross marks the current risk position $\bm{r}$ on the primary axes,
(ii) the dots mark the layer positions $\sigma_k$ on the context axes.
For analyzing a single context axis, it is particularly illustrative to keep
all other context axes constant and vary only one axis stepwise.
The dot on that axis then moves radially outward/inward, and changes in the
risk position (cross) become visible immediately.

\clearpage

\begin{figure*}[!t]
\centering

\subfloat[2D matrix slice $M_{\bm{\sigma}}$; Maintenance status = \texttt{recently serviced}.]{
\begin{tikzpicture}[scale=0.9, every node/.style={font=\footnotesize}]

  \def\W{16.8}

  \draw[rounded corners=1pt] (0,3.2) rectangle (\W,4.4);
  \node[anchor=west] at (0.3,4.15) {\textbf{Context layer (slice)}};

  \node[anchor=west] at (0.3,3.75) {Cooling redundancy level:};
  \draw[rounded corners=1pt] (4.1,3.55) rectangle (6.6,3.95);
  \node at (5.35,3.75) {$1{:}~N{+}1$};

  \node[anchor=west] at (7.2,3.75) {Maintenance status:};
  \draw[rounded corners=1pt] (10.2,3.55) rectangle (14.6,3.95); 
  \node at (12.4,3.75) {0: recently serviced};

  \draw (0,0) rectangle (\W,3.2);

  \node[anchor=west] at (0.3,3.00) {\textbf{2D-Risk Matrix} (Probability $\times$ Impact)};

  \def\xZero{7.0}
  \def\yZero{0.55}
  \def\cellW{1.55}
  \def\cellH{0.45}

  \definecolor{cG}{RGB}{0,230,0}
  \definecolor{cLG}{RGB}{160,240,0}
  \definecolor{cO}{RGB}{245,170,0}

  \foreach \i/\col in {0/cG,1/cG,2/cLG,3/cLG,4/cLG}{
    \fill[\col] ({\xZero+\i*\cellW},{\yZero+0*\cellH}) rectangle ++(\cellW,\cellH);
    \draw        ({\xZero+\i*\cellW},{\yZero+0*\cellH}) rectangle ++(\cellW,\cellH);
  }
  \foreach \i/\col in {0/cG,1/cLG,2/cLG,3/cLG,4/cO}{
    \fill[\col] ({\xZero+\i*\cellW},{\yZero+1*\cellH}) rectangle ++(\cellW,\cellH);
    \draw        ({\xZero+\i*\cellW},{\yZero+1*\cellH}) rectangle ++(\cellW,\cellH);
  }
  \foreach \i/\col in {0/cLG,1/cLG,2/cLG,3/cO,4/cO}{
    \fill[\col] ({\xZero+\i*\cellW},{\yZero+2*\cellH}) rectangle ++(\cellW,\cellH);
    \draw        ({\xZero+\i*\cellW},{\yZero+2*\cellH}) rectangle ++(\cellW,\cellH);
  }
  \foreach \i/\col in {0/cLG,1/cLG,2/cO,3/cO,4/cO}{
    \fill[\col] ({\xZero+\i*\cellW},{\yZero+3*\cellH}) rectangle ++(\cellW,\cellH);
    \draw        ({\xZero+\i*\cellW},{\yZero+3*\cellH}) rectangle ++(\cellW,\cellH);
  }
  \foreach \i/\col in {0/cLG,1/cO,2/cO,3/cO,4/cO}{
    \fill[\col] ({\xZero+\i*\cellW},{\yZero+4*\cellH}) rectangle ++(\cellW,\cellH);
    \draw        ({\xZero+\i*\cellW},{\yZero+4*\cellH}) rectangle ++(\cellW,\cellH);
  }

  \foreach \j/\t in {0/Insignificant,1/Low,2/Medium,3/High,4/Catastrophic}{
    \node[font=\scriptsize, anchor=east]
      at ({\xZero-0.25},{\yZero+(\j+0.5)*\cellH}) {\t};
  }

  \foreach \i/\t in {0/Very low,1/Low,2/Medium,3/High,4/Very high}{
    \node[font=\tiny]
      at ({\xZero+(\i+0.5)*\cellW},{\yZero-0.12}) {\t};
  }

  \node[font=\footnotesize, anchor=north]
    at ({\xZero+2.5*\cellW},{\yZero-0.19}) {Probability of occurrence};

  \node[font=\footnotesize, rotate=90]
    at ({\xZero-3.00},{\yZero+2.5*\cellH}) {Impact};

  \draw[black, line width=1.2pt]
    ({\xZero+2*\cellW},{\yZero+2*\cellH}) rectangle ++(\cellW,\cellH);

\end{tikzpicture}\label{fig:matrix_view_schematic}
}\hfill
\subfloat[ND polar heatmap: crosses mark probability/impact; dots mark the selected context levels.]{
\begin{tikzpicture}[scale=0.98, line cap=round, line join=round, every node/.style={font=\footnotesize}]

  \def\R{2.25}
  \def\start{-45}
  \def\step{90}

  \def\nProb{5}
  \def\nImp{5}
  \def\nCool{4}
  \def\nMaint{3}

  \definecolor{pG}{RGB}{88,241,89}
  \definecolor{pLG}{RGB}{190,241,89}
  \definecolor{pY}{RGB}{241,241,89}
  \definecolor{pO}{RGB}{241,190,89}
  \definecolor{pR}{RGB}{241,88,89}

  \def\fillwedge#1#2#3#4{%
    \path[fill=#4] (0,0)
      -- ({#3*cos(#1)},{#3*sin(#1)})
      arc[start angle=#1, end angle=#2, radius=#3] -- cycle;
  }
  \def\fillring#1#2#3#4#5{%
    \path[fill=#5] (0,0)
      -- ({#4*cos(#1)},{#4*sin(#1)})
      arc[start angle=#1, end angle=#2, radius=#4]
      -- ({#3*cos(#2)},{#3*sin(#2)})
      arc[start angle=#2, end angle=#1, radius=#3] -- cycle;
  }

  \foreach \q in {0,1,2,3}{
    \draw[black!22] (0,0) -- ({\R*cos(\start+\q*\step)},{\R*sin(\start+\q*\step)});
  }
  \draw[black!30] (0,0) circle (\R);

  \foreach \q in {0,1,2,3}{
    \pgfmathsetmacro{\angA}{\start+\q*\step}
    \pgfmathsetmacro{\angB}{\start+(\q+1)*\step}

    \def\nHere{5}\def\outerCol{pO}
    \ifnum\q=0 \def\nHere{\nImp}\def\outerCol{pO}\fi
    \ifnum\q=1 \def\nHere{\nProb}\def\outerCol{pO}\fi
    \ifnum\q=2 \def\nHere{\nMaint}\def\outerCol{pR}\fi
    \ifnum\q=3 \def\nHere{\nCool}\def\outerCol{pR}\fi
    \pgfmathtruncatemacro{\nTmp}{\nHere}

    \ifnum\nTmp=5
      \fillwedge{\angA}{\angB}{\R*1/\nTmp}{pG!85}
      \fillring{\angA}{\angB}{\R*1/\nTmp}{\R*2/\nTmp}{pLG!80}
      \fillring{\angA}{\angB}{\R*2/\nTmp}{\R*3/\nTmp}{pLG!80}
      \fillring{\angA}{\angB}{\R*3/\nTmp}{\R*4/\nTmp}{pO!100}
      \fillring{\angA}{\angB}{\R*4/\nTmp}{\R*5/\nTmp}{\outerCol!100}
      \foreach \k in {1,2,3,4}{
        \draw[black!18] ({\R*\k/\nTmp*cos(\angA)},{\R*\k/\nTmp*sin(\angA)})
          arc[start angle=\angA, end angle=\angB, radius=\R*\k/\nTmp];
      }
    \fi

    \ifnum\nTmp=4
      \fillwedge{\angA}{\angB}{\R*1/\nTmp}{pG!100}
      \fillring{\angA}{\angB}{\R*1/\nTmp}{\R*2/\nTmp}{pLG!80}
      \fillring{\angA}{\angB}{\R*2/\nTmp}{\R*3/\nTmp}{pO!100}
      \fillring{\angA}{\angB}{\R*3/\nTmp}{\R*4/\nTmp}{\outerCol!100}
      \foreach \k in {1,2,3}{
        \draw[black!18] ({\R*\k/\nTmp*cos(\angA)},{\R*\k/\nTmp*sin(\angA)})
          arc[start angle=\angA, end angle=\angB, radius=\R*\k/\nTmp];
      }
    \fi

    \ifnum\nTmp=3
      \fillwedge{\angA}{\angB}{\R*1/\nTmp}{pG!100}
      \fillring{\angA}{\angB}{\R*1/\nTmp}{\R*2/\nTmp}{pY!100}
      \fillring{\angA}{\angB}{\R*2/\nTmp}{\R*3/\nTmp}{\outerCol!100}
      \foreach \k in {1,2}{
        \draw[black!18] ({\R*\k/\nTmp*cos(\angA)},{\R*\k/\nTmp*sin(\angA)})
          arc[start angle=\angA, end angle=\angB, radius=\R*\k/\nTmp];
      }
    \fi
  }

  \draw[black!25] (-\R,0) -- (\R,0);
  \draw[black!25] (0,-\R) -- (0,\R);

  \draw[black, line width=1.2pt]
    ({\R*3/\nImp*cos(\start)},{\R*3/\nImp*sin(\start)})
    arc[start angle=\start, end angle=\start+\step, radius=\R*3/\nImp];

  \draw[black, line width=1.2pt]
    ({\R*4/\nProb*cos(\start+\step)},{\R*4/\nProb*sin(\start+\step)})
    arc[start angle=\start+\step, end angle=\start+2*\step, radius=\R*4/\nProb];

  \draw[black, line width=1.2pt]
    ({\R*2/\nMaint*cos(\start+2*\step)},{\R*2/\nMaint*sin(\start+2*\step)})
    arc[start angle=\start+2*\step, end angle=\start+3*\step, radius=\R*2/\nMaint];

  \draw[black, line width=1.2pt]
    ({\R*3/\nCool*cos(\start+3*\step)},{\R*3/\nCool*sin(\start+3*\step)})
    arc[start angle=\start+3*\step, end angle=\start+4*\step, radius=\R*3/\nCool];

  \def\kProb{3}   \def\labProb{Medium}
  \def\kImp{3}    \def\labImp{Medium}
  \def\kCool{2}   \def\labCool{$N{+}1$}
  \def\kMaint{1}  \def\labMaint{recently serviced}

  \pgfmathsetmacro{\rProb}{\R*(\kProb-0.5)/\nProb}
  \pgfmathsetmacro{\rImp}{\R*(\kImp-0.5)/\nImp}
  \pgfmathsetmacro{\rCool}{\R*(\kCool-0.5)/\nCool}
  \pgfmathsetmacro{\rMaint}{\R*(\kMaint-0.5)/\nMaint}

  \node[font=\large] at ({\rProb*cos(90)},{\rProb*sin(90)}) {$\times$};
  \node[font=\large] at ({\rImp*cos(0)},{\rImp*sin(0)}) {$\times$};

  \coordinate (pC) at ({\rCool*cos(270)},{\rCool*sin(270)});
  \coordinate (pM) at ({\rMaint*cos(180)},{\rMaint*sin(180)});
  \fill[black] (pC) circle (0.045);
  \fill[black] (pM) circle (0.045);
  \fill[black] (0,0) circle (0.03);

  \node[font=\scriptsize, anchor=south] at ({\rProb*cos(90)},{\rProb*sin(90)+0.10}) {\labProb};
  \node[font=\scriptsize, anchor=west]  at ({\rImp*cos(0)+0.12},{\rImp*sin(0)}) {\labImp};
  \node[font=\scriptsize, anchor=north] at ($(pC)+(0,-0.10)$) {\labCool};
  \node[font=\scriptsize, anchor=east]  at ($(pM)+(-0.12,0)$) {\labMaint};

  \node[align=center] at ({(\R+0.95)*cos(90)},{(\R+0.95)*sin(90)}) {Probability of\\occurrence};
  \node[align=center] at ({(\R+0.95)*cos(0)},{(\R+0.95)*sin(0)}) {Impact};
  \node[align=center] at ({(\R+1.05)*cos(270)},{(\R+1.05)*sin(270)}) {Cooling\\redundancy level};
  \node[align=center] at ({(\R+1.05)*cos(180)},{(\R+1.05)*sin(180)}) {Maintenance\\status};

\end{tikzpicture}\label{fig:polar_view_schematic}
}

\caption{Comparison: 2D risk matrix as a context-indexed slice $M_{\bm{\sigma}}$ vs.\ ND polar heatmap with axis-specific numbers of levels (5/5/4/3) and highlighting of the active layers.}
\label{fig:matrix_vs_polar_overview}
\end{figure*}
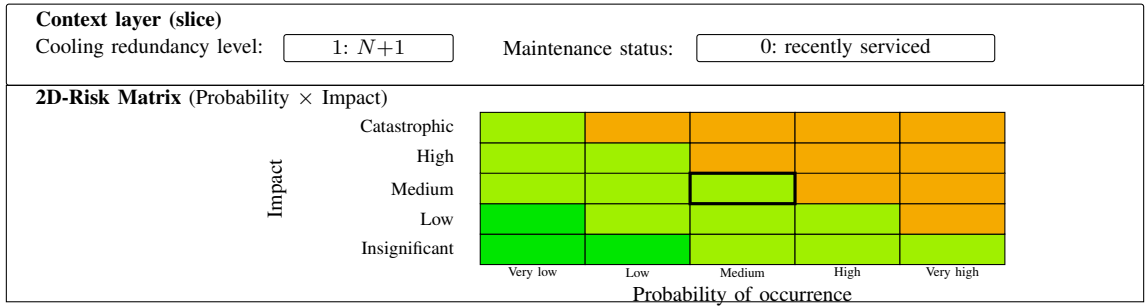
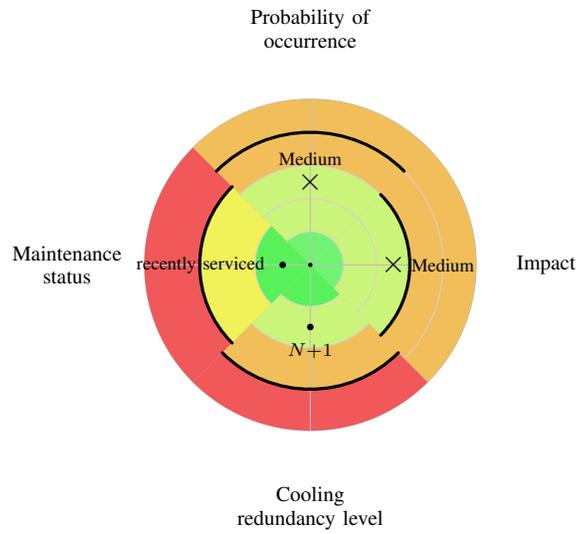

\begin{figure*}[!t]
\centering

\subfloat[2D matrix slice $M_{\bm{\sigma}}$; Maintenance status = \texttt{due}.]{
\begin{tikzpicture}[scale=0.9, every node/.style={font=\footnotesize}]

  \def\W{16.8}

  \draw[rounded corners=1pt] (0,3.2) rectangle (\W,4.4);
  \node[anchor=west] at (0.3,4.15) {\textbf{Context layer (slice)}};

  \node[anchor=west] at (0.3,3.75) {Cooling redundancy level:};
  \draw[rounded corners=1pt] (4.1,3.55) rectangle (6.6,3.95);
  \node at (5.35,3.75) {$1{:}~N{+}1$};

  \node[anchor=west] at (7.2,3.75) {Maintenance status:};
  \draw[rounded corners=1pt] (10.2,3.55) rectangle (14.6,3.95);
  \node at (12.4,3.75) {1: due};

  \draw (0,0) rectangle (\W,3.2);
  \node[anchor=west] at (0.3,3.05) {\textbf{2D-Risk Matrix} (Probability $\times$ Impact)};

  \def\xZero{7.0}
  \def\yZero{0.60}
  \def\cellW{1.55}
  \def\cellH{0.45}

  \definecolor{cG}{RGB}{0,230,0}
  \definecolor{cLG}{RGB}{160,240,0}
  \definecolor{cO}{RGB}{245,170,0}
  \definecolor{cR}{RGB}{235,0,0}

  \foreach \i/\col in {0/cG,1/cLG,2/cLG,3/cLG,4/cO}{
    \fill[\col] ({\xZero+\i*\cellW},{\yZero+0*\cellH}) rectangle ++(\cellW,\cellH);
    \draw        ({\xZero+\i*\cellW},{\yZero+0*\cellH}) rectangle ++(\cellW,\cellH);
  }
  \foreach \i/\col in {0/cLG,1/cLG,2/cLG,3/cO,4/cO}{
    \fill[\col] ({\xZero+\i*\cellW},{\yZero+1*\cellH}) rectangle ++(\cellW,\cellH);
    \draw        ({\xZero+\i*\cellW},{\yZero+1*\cellH}) rectangle ++(\cellW,\cellH);
  }
  \foreach \i/\col in {0/cLG,1/cLG,2/cO,3/cO,4/cO}{
    \fill[\col] ({\xZero+\i*\cellW},{\yZero+2*\cellH}) rectangle ++(\cellW,\cellH);
    \draw        ({\xZero+\i*\cellW},{\yZero+2*\cellH}) rectangle ++(\cellW,\cellH);
  }
  \foreach \i/\col in {0/cLG,1/cO,2/cO,3/cO,4/cO}{
    \fill[\col] ({\xZero+\i*\cellW},{\yZero+3*\cellH}) rectangle ++(\cellW,\cellH);
    \draw        ({\xZero+\i*\cellW},{\yZero+3*\cellH}) rectangle ++(\cellW,\cellH);
  }
  \foreach \i/\col in {0/cO,1/cO,2/cO,3/cO,4/cR}{
    \fill[\col] ({\xZero+\i*\cellW},{\yZero+4*\cellH}) rectangle ++(\cellW,\cellH);
    \draw        ({\xZero+\i*\cellW},{\yZero+4*\cellH}) rectangle ++(\cellW,\cellH);
  }

  \foreach \j/\t in {0/Insignificant,1/Low,2/Medium,3/High,4/Catastrophic}{
    \node[font=\scriptsize, anchor=east] at ({\xZero-0.25},{\yZero+(\j+0.5)*\cellH}) {\t};
  }

  \foreach \i/\t in {0/Very low,1/Low,2/Medium,3/High,4/Very high}{
    \node[font=\tiny] at ({\xZero+(\i+0.5)*\cellW},{\yZero-0.08}) {\t};
  }

  \node[font=\footnotesize, anchor=north]
    at ({\xZero+2.5*\cellW},{\yZero-0.19}) {Probability of occurrence};

  \node[font=\footnotesize, rotate=90]
    at ({\xZero-3.00},{\yZero+2.5*\cellH}) {Impact};

  \draw[black, line width=1.2pt]
    ({\xZero+2*\cellW},{\yZero+2*\cellH}) rectangle ++(\cellW,\cellH);

\end{tikzpicture}\label{fig:matrix_view_due}
}\hfill
\subfloat[ND-Polar-Heatmap; Maintenance status = \texttt{due}.]{
\begin{tikzpicture}[scale=0.98, line cap=round, line join=round, every node/.style={font=\footnotesize}]

  \def\R{2.25}
  \def\start{-45}
  \def\step{90}

  \def\nProb{5}
  \def\nImp{5}
  \def\nCool{4}
  \def\nMaint{3}

  \definecolor{pG}{RGB}{88,241,89}
  \definecolor{pLG}{RGB}{190,241,89}
  \definecolor{pY}{RGB}{241,241,89}
  \definecolor{pO}{RGB}{241,190,89}
  \definecolor{pR}{RGB}{241,88,89}

  \def\fillwedge#1#2#3#4{%
    \path[fill=#4] (0,0)
      -- ({#3*cos(#1)},{#3*sin(#1)})
      arc[start angle=#1, end angle=#2, radius=#3] -- cycle;
  }
  \def\fillring#1#2#3#4#5{%
    \path[fill=#5] (0,0)
      -- ({#4*cos(#1)},{#4*sin(#1)})
      arc[start angle=#1, end angle=#2, radius=#4]
      -- ({#3*cos(#2)},{#3*sin(#2)})
      arc[start angle=#2, end angle=#1, radius=#3] -- cycle;
  }

  \foreach \q in {0,1,2,3}{
    \draw[black!22] (0,0) -- ({\R*cos(\start+\q*\step)},{\R*sin(\start+\q*\step)});
  }
  \draw[black!30] (0,0) circle (\R);

  \foreach \q in {0,1,2,3}{
    \pgfmathsetmacro{\angA}{\start+\q*\step}
    \pgfmathsetmacro{\angB}{\start+(\q+1)*\step}

    \def\nHere{5}\def\outerCol{pO}
    \ifnum\q=0 \def\nHere{\nImp}\def\outerCol{pO}\fi
    \ifnum\q=1 \def\nHere{\nProb}\def\outerCol{pO}\fi
    \ifnum\q=2 \def\nHere{\nMaint}\def\outerCol{pR}\fi
    \ifnum\q=3 \def\nHere{\nCool}\def\outerCol{pR}\fi
    \pgfmathtruncatemacro{\nTmp}{\nHere}

\ifnum\nTmp=5
  \fillwedge{\angA}{\angB}{\R*1/\nTmp}{pLG!80}          
  \fillring{\angA}{\angB}{\R*1/\nTmp}{\R*2/\nTmp}{pLG!80}
  \fillring{\angA}{\angB}{\R*2/\nTmp}{\R*3/\nTmp}{pO!100} 
  \fillring{\angA}{\angB}{\R*3/\nTmp}{\R*4/\nTmp}{pO!100}
  \fillring{\angA}{\angB}{\R*4/\nTmp}{\R*5/\nTmp}{\outerCol!100}
  \foreach \k in {1,2,3,4}{
    \draw[black!18] ({\R*\k/\nTmp*cos(\angA)},{\R*\k/\nTmp*sin(\angA)})
      arc[start angle=\angA, end angle=\angB, radius=\R*\k/\nTmp];
  }
\fi

    \ifnum\nTmp=4
      \fillwedge{\angA}{\angB}{\R*1/\nTmp}{pG!100}
      \fillring{\angA}{\angB}{\R*1/\nTmp}{\R*2/\nTmp}{pLG!80}
      \fillring{\angA}{\angB}{\R*2/\nTmp}{\R*3/\nTmp}{pO!100}
      \fillring{\angA}{\angB}{\R*3/\nTmp}{\R*4/\nTmp}{\outerCol!100}
      \foreach \k in {1,2,3}{
        \draw[black!18] ({\R*\k/\nTmp*cos(\angA)},{\R*\k/\nTmp*sin(\angA)})
          arc[start angle=\angA, end angle=\angB, radius=\R*\k/\nTmp];
      }
    \fi

    \ifnum\nTmp=3
      \fillwedge{\angA}{\angB}{\R*1/\nTmp}{pG!100}
      \fillring{\angA}{\angB}{\R*1/\nTmp}{\R*2/\nTmp}{pY!100}
      \fillring{\angA}{\angB}{\R*2/\nTmp}{\R*3/\nTmp}{\outerCol!100}
      \foreach \k in {1,2}{
        \draw[black!18] ({\R*\k/\nTmp*cos(\angA)},{\R*\k/\nTmp*sin(\angA)})
          arc[start angle=\angA, end angle=\angB, radius=\R*\k/\nTmp];
      }
    \fi
  }

  \draw[black!25] (-\R,0) -- (\R,0);
  \draw[black!25] (0,-\R) -- (0,\R);

  \draw[black, line width=1.2pt]
    ({\R*3/\nImp*cos(\start)},{\R*3/\nImp*sin(\start)})
    arc[start angle=\start, end angle=\start+\step, radius=\R*3/\nImp];
  \draw[black, line width=1.2pt]
    ({\R*4/\nProb*cos(\start+\step)},{\R*4/\nProb*sin(\start+\step)})
    arc[start angle=\start+\step, end angle=\start+2*\step, radius=\R*4/\nProb];
  \draw[black, line width=1.2pt]
    ({\R*2/\nMaint*cos(\start+2*\step)},{\R*2/\nMaint*sin(\start+2*\step)})
    arc[start angle=\start+2*\step, end angle=\start+3*\step, radius=\R*2/\nMaint];
  \draw[black, line width=1.2pt]
    ({\R*3/\nCool*cos(\start+3*\step)},{\R*3/\nCool*sin(\start+3*\step)})
    arc[start angle=\start+3*\step, end angle=\start+4*\step, radius=\R*3/\nCool];

  \def\kProb{3}   \def\labProb{Medium}
  \def\kImp{3}    \def\labImp{Medium}
  \def\kCool{2}   \def\labCool{$N{+}1$}
  \def\kMaint{2}  \def\labMaint{due} 

  \pgfmathsetmacro{\rProb}{\R*(\kProb-0.5)/\nProb}
  \pgfmathsetmacro{\rImp}{\R*(\kImp-0.5)/\nImp}
  \pgfmathsetmacro{\rCool}{\R*(\kCool-0.5)/\nCool}
  \pgfmathsetmacro{\rMaint}{\R*(\kMaint-0.5)/\nMaint}

  \node[font=\large] at ({\rProb*cos(90)},{\rProb*sin(90)}) {$\times$};
  \node[font=\large] at ({\rImp*cos(0)},{\rImp*sin(0)}) {$\times$};

  \coordinate (pC) at ({\rCool*cos(270)},{\rCool*sin(270)});
  \coordinate (pM) at ({\rMaint*cos(180)},{\rMaint*sin(180)});
  \fill[black] (pC) circle (0.045);
  \fill[black] (pM) circle (0.045);
  \fill[black] (0,0) circle (0.03);

  \node[font=\scriptsize, anchor=south] at ({\rProb*cos(90)},{\rProb*sin(90)+0.10}) {\labProb};
  \node[font=\scriptsize, anchor=west]  at ({\rImp*cos(0)+0.12},{\rImp*sin(0)}) {\labImp};
  \node[font=\scriptsize, anchor=north] at ($(pC)+(0,-0.10)$) {\labCool};
  \node[font=\scriptsize, anchor=east]  at ($(pM)+(-0.12,0)$) {\labMaint};

  \node[align=center] at ({(\R+0.95)*cos(90)},{(\R+0.95)*sin(90)}) {Probability of\\occurrence};
  \node[align=center] at ({(\R+0.95)*cos(0)},{(\R+0.95)*sin(0)}) {Impact};
  \node[align=center] at ({(\R+1.05)*cos(270)},{(\R+1.05)*sin(270)}) {Cooling\\redundancy level};
  \node[align=center] at ({(\R+1.05)*cos(180)},{(\R+1.05)*sin(180)}) {Maintenance\\status};

\end{tikzpicture}\label{fig:polar_view_due}
}

\caption{Comparison (Maintenance = \texttt{due}): 2D slice $M_{\bm{\sigma}}$ with adjusted matrix colouring vs.\ ND polar heatmap with highlighted active layers.}
\label{fig:matrix_vs_polar_due}
\end{figure*}
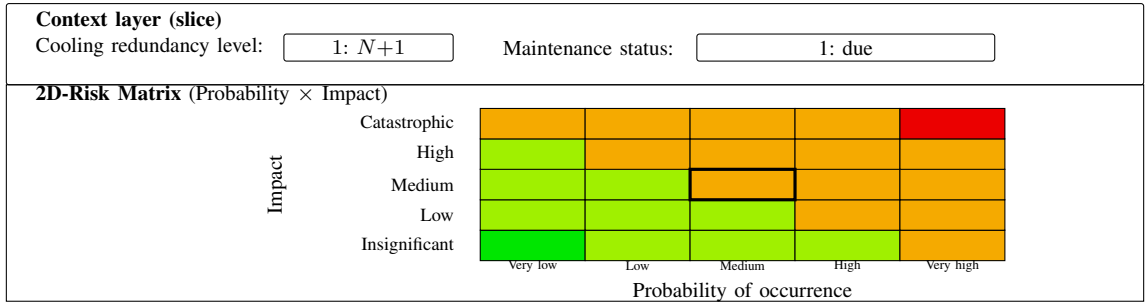
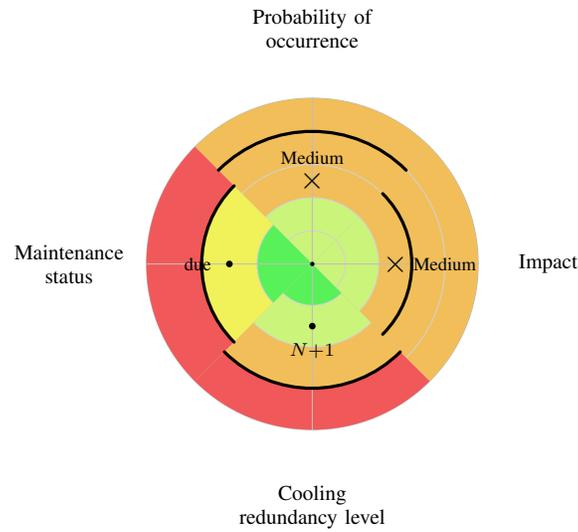

\begin{figure*}[!t]
\centering

\subfloat[2D matrix slice $M_{\bm{\sigma}}$; Maintenance status = \texttt{overdue}.]{
\begin{tikzpicture}[scale=0.9, every node/.style={font=\footnotesize}]

  \def\W{16.8}

  \draw[rounded corners=1pt] (0,3.2) rectangle (\W,4.4);
  \node[anchor=west] at (0.3,4.15) {\textbf{Context layer (slice)}};

  \node[anchor=west] at (0.3,3.75) {Cooling redundancy level:};
  \draw[rounded corners=1pt] (4.1,3.55) rectangle (6.6,3.95);
  \node at (5.35,3.75) {$1{:}~N{+}1$};

  \node[anchor=west] at (7.2,3.75) {Maintenance status:};
  \draw[rounded corners=1pt] (10.2,3.55) rectangle (14.6,3.95);
  \node at (12.4,3.75) {2: overdue};

  \draw (0,0) rectangle (\W,3.2);
  \node[anchor=west] at (0.3,3.05) {\textbf{2D-Risk Matrix} (Probability $\times$ Impact)};

  \def\xZero{7.0}
  \def\yZero{0.60}
  \def\cellW{1.55}
  \def\cellH{0.45}

  \definecolor{cG}{RGB}{0,230,0}
  \definecolor{cLG}{RGB}{160,240,0}
  \definecolor{cO}{RGB}{245,170,0}
  \definecolor{cR}{RGB}{235,0,0}

  \foreach \i/\col in {0/cG,1/cG,2/cLG,3/cLG,4/cO}{
    \fill[\col] ({\xZero+\i*\cellW},{\yZero+0*\cellH}) rectangle ++(\cellW,\cellH);
    \draw        ({\xZero+\i*\cellW},{\yZero+0*\cellH}) rectangle ++(\cellW,\cellH);
  }
  \foreach \i/\col in {0/cG,1/cLG,2/cLG,3/cO,4/cO}{
    \fill[\col] ({\xZero+\i*\cellW},{\yZero+1*\cellH}) rectangle ++(\cellW,\cellH);
    \draw        ({\xZero+\i*\cellW},{\yZero+1*\cellH}) rectangle ++(\cellW,\cellH);
  }
  \foreach \i/\col in {0/cLG,1/cLG,2/cO,3/cO,4/cO}{
    \fill[\col] ({\xZero+\i*\cellW},{\yZero+2*\cellH}) rectangle ++(\cellW,\cellH);
    \draw        ({\xZero+\i*\cellW},{\yZero+2*\cellH}) rectangle ++(\cellW,\cellH);
  }
  \foreach \i/\col in {0/cLG,1/cO,2/cO,3/cO,4/cR}{
    \fill[\col] ({\xZero+\i*\cellW},{\yZero+3*\cellH}) rectangle ++(\cellW,\cellH);
    \draw        ({\xZero+\i*\cellW},{\yZero+3*\cellH}) rectangle ++(\cellW,\cellH);
  }
  \foreach \i/\col in {0/cO,1/cO,2/cO,3/cR,4/cR}{
    \fill[\col] ({\xZero+\i*\cellW},{\yZero+4*\cellH}) rectangle ++(\cellW,\cellH);
    \draw        ({\xZero+\i*\cellW},{\yZero+4*\cellH}) rectangle ++(\cellW,\cellH);
  }

  \foreach \j/\t in {0/Insignificant,1/Low,2/Medium,3/High,4/Catastrophic}{
    \node[font=\scriptsize, anchor=east] at ({\xZero-0.25},{\yZero+(\j+0.5)*\cellH}) {\t};
  }

  \foreach \i/\t in {0/Very low,1/Low,2/Medium,3/High,4/Very high}{
    \node[font=\tiny] at ({\xZero+(\i+0.5)*\cellW},{\yZero-0.08}) {\t};
  }

  \node[font=\footnotesize, anchor=north]
    at ({\xZero+2.5*\cellW},{\yZero-0.19}) {Probability of occurrence};

  \node[font=\footnotesize, rotate=90]
    at ({\xZero-3.00},{\yZero+2.5*\cellH}) {Impact};

  \draw[black, line width=1.2pt]
    ({\xZero+2*\cellW},{\yZero+2*\cellH}) rectangle ++(\cellW,\cellH);

\end{tikzpicture}\label{fig:matrix_view_overdue}
}\hfill
\subfloat[ND-Polar-Heatmap; Maintenance status = \texttt{overdue}.]{
\begin{tikzpicture}[scale=0.98, line cap=round, line join=round, every node/.style={font=\footnotesize}]

  \def\R{2.25}
  \def\start{-45}
  \def\step{90}

  \def\nProb{5}
  \def\nImp{5}
  \def\nCool{4}
  \def\nMaint{3}

  \definecolor{pG}{RGB}{88,241,89}
  \definecolor{pLG}{RGB}{190,241,89}
  \definecolor{pY}{RGB}{241,241,89}
  \definecolor{pO}{RGB}{241,190,89}
  \definecolor{pR}{RGB}{241,88,89}

  \def\fillwedge#1#2#3#4{%
    \path[fill=#4] (0,0)
      -- ({#3*cos(#1)},{#3*sin(#1)})
      arc[start angle=#1, end angle=#2, radius=#3] -- cycle;
  }
  \def\fillring#1#2#3#4#5{%
    \path[fill=#5] (0,0)
      -- ({#4*cos(#1)},{#4*sin(#1)})
      arc[start angle=#1, end angle=#2, radius=#4]
      -- ({#3*cos(#2)},{#3*sin(#2)})
      arc[start angle=#2, end angle=#1, radius=#3] -- cycle;
  }

  \foreach \q in {0,1,2,3}{
    \draw[black!22] (0,0) -- ({\R*cos(\start+\q*\step)},{\R*sin(\start+\q*\step)});
  }
  \draw[black!30] (0,0) circle (\R);

  \foreach \q in {0,1,2,3}{
    \pgfmathsetmacro{\angA}{\start+\q*\step}
    \pgfmathsetmacro{\angB}{\start+(\q+1)*\step}

    \def\nHere{5}\def\outerCol{pO}
    \ifnum\q=0 \def\nHere{\nImp}\def\outerCol{pO}\fi
    \ifnum\q=1 \def\nHere{\nProb}\def\outerCol{pO}\fi
    \ifnum\q=2 \def\nHere{\nMaint}\def\outerCol{pR}\fi
    \ifnum\q=3 \def\nHere{\nCool}\def\outerCol{pR}\fi
    \pgfmathtruncatemacro{\nTmp}{\nHere}

    \ifnum\nTmp=5
      \fillwedge{\angA}{\angB}{\R*1/\nTmp}{pLG!80}
      \fillring{\angA}{\angB}{\R*1/\nTmp}{\R*2/\nTmp}{pLG!80}
      \fillring{\angA}{\angB}{\R*2/\nTmp}{\R*3/\nTmp}{pO!100}
      \fillring{\angA}{\angB}{\R*3/\nTmp}{\R*4/\nTmp}{pO!100}
      \fillring{\angA}{\angB}{\R*4/\nTmp}{\R*5/\nTmp}{\outerCol!100}
      \foreach \k in {1,2,3,4}{
        \draw[black!18] ({\R*\k/\nTmp*cos(\angA)},{\R*\k/\nTmp*sin(\angA)})
          arc[start angle=\angA, end angle=\angB, radius=\R*\k/\nTmp];
      }
    \fi

    \ifnum\nTmp=4
      \fillwedge{\angA}{\angB}{\R*1/\nTmp}{pG!100}
      \fillring{\angA}{\angB}{\R*1/\nTmp}{\R*2/\nTmp}{pLG!80}
      \fillring{\angA}{\angB}{\R*2/\nTmp}{\R*3/\nTmp}{pO!100}
      \fillring{\angA}{\angB}{\R*3/\nTmp}{\R*4/\nTmp}{\outerCol!100}
      \foreach \k in {1,2,3}{
        \draw[black!18] ({\R*\k/\nTmp*cos(\angA)},{\R*\k/\nTmp*sin(\angA)})
          arc[start angle=\angA, end angle=\angB, radius=\R*\k/\nTmp];
      }
    \fi

    \ifnum\nTmp=3
      \fillwedge{\angA}{\angB}{\R*1/\nTmp}{pG!100}
      \fillring{\angA}{\angB}{\R*1/\nTmp}{\R*2/\nTmp}{pY!100}
      \fillring{\angA}{\angB}{\R*2/\nTmp}{\R*3/\nTmp}{\outerCol!100}
      \foreach \k in {1,2}{
        \draw[black!18] ({\R*\k/\nTmp*cos(\angA)},{\R*\k/\nTmp*sin(\angA)})
          arc[start angle=\angA, end angle=\angB, radius=\R*\k/\nTmp];
      }
    \fi
  }

  \draw[black!25] (-\R,0) -- (\R,0);
  \draw[black!25] (0,-\R) -- (0,\R);

  \draw[black, line width=1.2pt]
    ({\R*3/\nImp*cos(\start)},{\R*3/\nImp*sin(\start)})
    arc[start angle=\start, end angle=\start+\step, radius=\R*3/\nImp];
  \draw[black, line width=1.2pt]
    ({\R*4/\nProb*cos(\start+\step)},{\R*4/\nProb*sin(\start+\step)})
    arc[start angle=\start+\step, end angle=\start+2*\step, radius=\R*4/\nProb];
  \draw[black, line width=1.2pt]
    ({\R*2/\nMaint*cos(\start+2*\step)},{\R*2/\nMaint*sin(\start+2*\step)})
    arc[start angle=\start+2*\step, end angle=\start+3*\step, radius=\R*2/\nMaint];
  \draw[black, line width=1.2pt]
    ({\R*3/\nCool*cos(\start+3*\step)},{\R*3/\nCool*sin(\start+3*\step)})
    arc[start angle=\start+3*\step, end angle=\start+4*\step, radius=\R*3/\nCool];

  \def\kProb{3}   \def\labProb{Medium}
  \def\kImp{3}    \def\labImp{Medium}
  \def\kCool{2}   \def\labCool{$N{+}1$}
  \def\kMaint{3}  \def\labMaint{overdue} 

  \pgfmathsetmacro{\rProb}{\R*(\kProb-0.5)/\nProb}
  \pgfmathsetmacro{\rImp}{\R*(\kImp-0.5)/\nImp}
  \pgfmathsetmacro{\rCool}{\R*(\kCool-0.5)/\nCool}
  \pgfmathsetmacro{\rMaint}{\R*(\kMaint-0.5)/\nMaint}

  \node[font=\large] at ({\rProb*cos(90)},{\rProb*sin(90)}) {$\times$};
  \node[font=\large] at ({\rImp*cos(0)},{\rImp*sin(0)}) {$\times$};

  \coordinate (pC) at ({\rCool*cos(270)},{\rCool*sin(270)});
  \coordinate (pM) at ({\rMaint*cos(180)},{\rMaint*sin(180)});
  \fill[black] (pC) circle (0.045);
  \fill[black] (pM) circle (0.045);
  \fill[black] (0,0) circle (0.03);

  \node[font=\scriptsize, anchor=south] at ({\rProb*cos(90)},{\rProb*sin(90)+0.10}) {\labProb};
  \node[font=\scriptsize, anchor=west]  at ({\rImp*cos(0)+0.12},{\rImp*sin(0)}) {\labImp};
  \node[font=\scriptsize, anchor=north] at ($(pC)+(0,-0.10)$) {\labCool};
  \node[font=\scriptsize, anchor=east]  at ($(pM)+(-0.12,0)$) {\labMaint};

  \node[align=center] at ({(\R+0.95)*cos(90)},{(\R+0.95)*sin(90)}) {Probability of\\occurrence};
  \node[align=center] at ({(\R+0.95)*cos(0)},{(\R+0.95)*sin(0)}) {Impact};
  \node[align=center] at ({(\R+1.05)*cos(270)},{(\R+1.05)*sin(270)}) {Cooling\\redundancy level};
  \node[align=center] at ({(\R+1.05)*cos(180)},{(\R+1.05)*sin(180)}) {Maintenance\\status};

\end{tikzpicture}\label{fig:polar_view_overdue}
}

\caption{Comparison (Maintenance = \texttt{overdue}): 2D slice $M_{\bm{\sigma}}$ with UI-conform matrix colouring (incl.\ red) vs.\ ND polar heatmap with highlighted active layers.}
\label{fig:matrix_vs_polar_overdue}
\end{figure*}
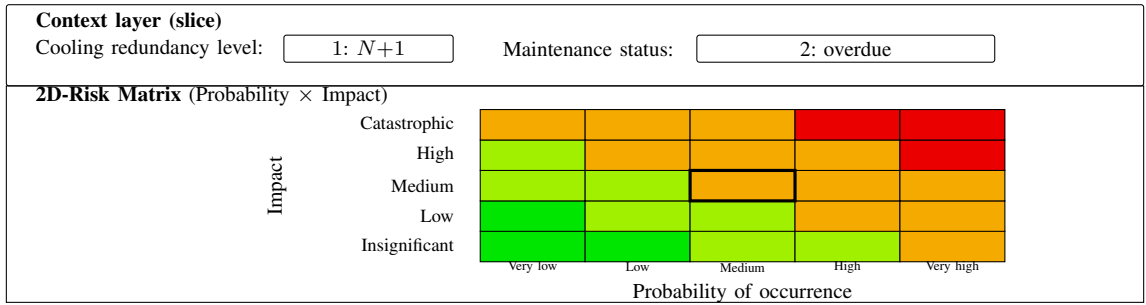
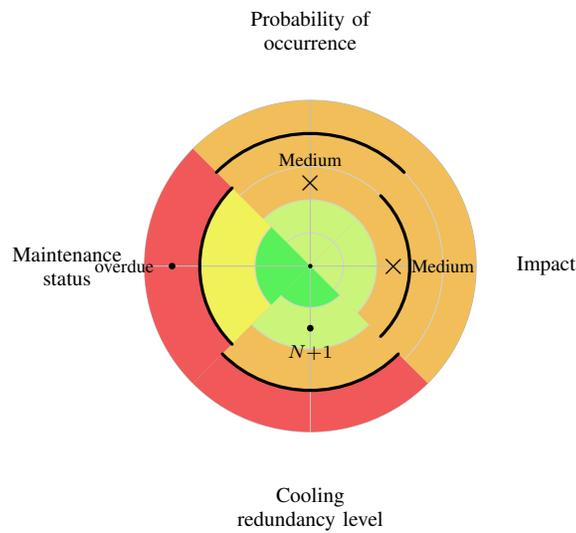
\clearpage

\subsection{Observation in the cooling scenario: varying redundancy with constant maintenance status}
\label{sec:cooling_redundancy_walk}

Fig.~\ref{fig:matrix_vs_polar_overview} schematically shows two visualizations. The 2D view (a) of the current slice is given as for the context state, where cooling redundancy level = $N+1$ and maintenance status = \textit{recently serviced}.
The polar view (b) for the same context state is shown below. 

In the same visualization style, Fig.~\ref{fig:matrix_vs_polar_due} and Fig.~\ref{fig:matrix_vs_polar_overdue}
illustrate the stepwise reduction of the maintenance status (e.\,g.\
\emph{recently serviced} $\rightarrow$ \emph{due} $\rightarrow$ \emph{overdue}) while the
redunancy level ($N+1$) is held constant.
The representation highlights that the polar heatmap supports \emph{exploration}
along a single context dimension with the other dimensions held constant as a
continuous visual trajectory, whereas the 2D matrix requires discrete switching
between multiple slices.

\section{Conclusion}
Multidimensional polar heatmaps provide a formally consistent and visually
interpretable approach to extending classical heatmap-based risk analyses.
They enable the explicit integration of context factors and thereby address
central requirements of current resilience and regulatory frameworks such as NIS2 and DORA.

As an outlook, we will advance this line of work through a series of developments towards risk management for complex infrastructures and systems under the umbrella term \textit{Hagenberg Risk Management Process}. The ND multidimensional polar heatmap introduced in this paper constitutes a first step, providing a formal yet practical basis for integrating contextual dimensions into risk modelling and triage.

To support future case studies and further research, ND multidimensional polar heatmaps have already been implemented in the real-time risk management solution \emph{Riskomat}. For further information about the solution, readers may contact the authors via email at \emph{info@kompilomat.com}.

\bibliographystyle{IEEEtran}

\end{document}